\documentclass[useAMS,usenatbib]{mn2e}

\usepackage{epsfig,times,graphicx,bm,amssymb,aas_macros}
\newcommand{\mk}{}
\newcommand{\rd}{{\rm d}}

\newcommand{\beq}{\begin{equation}}
\newcommand{\eeq}{\end{equation}}
\newcommand{\beqa}{\begin{eqnarray}}
\newcommand{\eeqa}{\end{eqnarray}}
\newcommand{\bcom}{}

\title[On the magnetic flux problem in star formation]{On the magnetic flux problem in star formation}
\author[Jonathan Braithwaite]{Jonathan Braithwaite \thanks{E-mail: jonathan@astro.uni-bonn.de}\\Argelander Institut f\"ur Astronomie, Auf dem H\"ugel 71, 53121 Bonn, Germany}
\pagerange{\pageref{firstpage}--\pageref{lastpage}}
\begin{document}\maketitle\label{firstpage}
\begin{abstract} Strong magnetic fields play a crucial role in the removal of angular momentum from collapsing clouds and protostellar discs and are necessary for the formation of disc winds as well as jets from the inner disc 
 and indeed, strong large-scale poloidal magnetic fields are observed in protostellar discs at all radii down to $\sim 10 R_\odot$. Nevertheless, by the time the star is visible virtually all of the original magnetic flux has vanished. I explore mechanisms for removing this flux during the formation of the protostar once it is magnetically disconnected from the parent cloud, looking at both radiative and convective protostars. This includes a numerical investigation of buoyant magnetic field removal from convective stars. It is found that if the star goes through a fully convective phase all remaining flux can easily be removed from the protostar, essentially on an Alfv\'en timescale. If on the other hand the protostar has no fully convective phase then some flux can be retained, the quantity depending on the net magnetic helicity, which is probably quite small. Only some fraction of this flux is visible at the stellar surface. I also look at how the same mechanisms could prevent flux from accreting onto the star at all, meaning that mass would only accrete as fast as it is able to slip past the flux.\end{abstract}
\begin{keywords} ({\it magnetohydrodynamics}) MHD -- stars: magnetic fields -- ISM: clouds -- ISM: magnetic fields -- stars: formation -- stars: magnetic fields \end{keywords}

\section{Introduction}
\label{sec:intro}

Stars form from clouds in which gravitational, thermal, {\mk rotational and magnetic energies are generally comparable to one another. Once a star has formed, the thermal and gravitational energies are still comparable to each other but the magnetic energy} (as inferred from Zeeman measurements) is at least six orders of magnitude lower in even the most strongly magnetised stars.

{\mk This is not what one {\it prima facie} expects if a cloud retains its magnetic flux as it collapses, since gravitational and magnetic energy have the same scaling with radius, i.e. $E\propto1/R$. In contrast, thermal and rotational energy rise faster, and must therefore be continually extracted from the cloud to allow its collapse. Extraction of thermal energy happens via radiation and represents no barrier to star formation in theory or otherwise (at least for stars under about $10M_\odot$); extraction of rotational energy is less well understood but theoretical mechanisms exist, using magnetic fields or gravitational instability, and there is plenty of evidence} for magnetic braking (\citealt{Gillis_etal:1974,Gillis_etal:1979,Mouschovias_Paleologou:1979,Stahler_Palla:2005} and refs.\ therein).

The relative strength of the gravitational and magnetic fields is often expressed as a dimensionless mass-to-flux ratio, defined as $\lambda\equiv 2\pi G^{1/2} M/\Phi$, {\mk where $M$ and $\Phi$ are the mass and magnetic flux,} or locally in a disc context as $2\pi G^{1/2}\Sigma/B_z$ where $B_z$ and $\Sigma$ are the field normal to the disc, and surface density; {\mk this ratio is conserved if flux freezing is valid. It is related to the gravitational and magnetic energies by (ignoring factors of order unity) $\lambda^2 = |E_{\rm grav}|/E_{\rm mag}$.} A cloud with $\lambda\gtrsim1$ is said to be `magnetically supercritical' and will collapse, in the absence of significant thermal or rotational energy. Conversely a cloud with $\lambda\lesssim1$ is `magnetically subcritical' and the magnetic field supports the cloud against gravity.

Observations of cloud cores and the ISM (e.g.\ \citealt{Crutcher_etal:1999,Heiles_Troland:2005,Girart:2006,Girart:2009}) show that clouds do contain strong magnetic fields and appear to be mildly magnetically subcritical on scales above $\sim 1000$ AU. The rotational and thermal energies appear to be relatively small on these large scales. 
 The so-called ``magnetic flux problem'' can be divided into two parts, the first of which is: how is sufficient magnetic flux lost so that the cloud becomes magnetically supercritical? Gravitational contraction down to that scale could proceed via ambipolar diffusion \citep{Mestel_Spitzer:1956} {\mk or some other diffusive process, and dynamical collapse begins once supercriticality is reached.} Recent observational support for this model is given by \citet{Davidson:2011}.

The second part of the magnetic flux problem -- that addressed here -- is how to remove the remaining flux once the cloud has become supercritical, to explain the incredibly weak magnetic fields seen in stars {\mk (e.g. \citealt{Mestel_Spitzer:1956,Nakano:1984,Galli:2009}).} The supercritical collapse might take place too quickly to allow ambipolar diffusion to remove the rest of the flux before flux freezing resumes owing to rising ionisation fraction (although opinions differ on this point, see \citet{Li:1998,Desch_Mouschovias:2001,Banerjee_Pudritz:2006} amongst others). 

In this paper I look at a solution to this problem, namely that the collapse can proceed with as little or as much loss of flux as dissipative processes allow, and that the excess flux is destroyed once the protostar forms, or rather, once it has become disconnected in some sense from its parent cloud. In the next section I review relevant observational and theoretical results of collapse from $1000$ AU to protostar formation. In section~\ref{sec:destruction}, I explore the mechanisms for destroying flux in stars via magnetohydrodynamic (MHD) instability in non-convective stars and buoyancy in convective stars, before investigating the latter numerically in section~\ref{sec:numerical}. I
 discuss the results in section~\ref{sec:discussion} and conclude in section~\ref{sec:conc}.

\section{Accretion and protostar formation}

I now summarise relevant results and evidence that at least part of the solution to the magnetic flux problem must lie in, or in the immediate vicinity of, the protostar.

\subsection{Inwards advection of magnetic flux in a disc}\label{sec:disc}

Once a cloud has become supercritical, it can collapse dynamically. Magnetic braking becomes ineffective once the collapse is super-Alfv\'enic, so the rotational energy becomes larger in relation to the other energies. Normally this leads to formation of a disc of radius $100-1000$ AU, although some systems might lack a disc \citep{Stassun_etal:1999,Stassun_etal:2001,Rebull_etal:2006}. 
 However, we shall look here at the magnetic flux accreted via a disc. 
 
Gas in an accretion disc can lose its angular momentum and spiral inwards by passing it either vertically to a disc wind or jet \citep{Blandford_Payne:1982}, or radially outwards to disc gas exterior to itself via some instability-driven turbulence. The instability could be gravitational (e.g.\ \citealt{Jappsen_Klessen:2004,Vorobyov_Basu:2006,Zhu_etal:2009}), magnetorotational (MRI, \citealt{Balbus:1991,Pessah:2010} and refs.\ therein) perhaps with buoyancy (e.g.\ \citealt{Keppens_etal:2002}) or purely hydrodynamical \citep{Lithwick:2009}.

Jets are observed in many objects, are launched from radii $\sim1$ AU \citep{Bacciotti_etal:2002,Anderson_etal:2003} and often appear together with a slower, less collimated wind. Around a tenth of the accreted mass is ejected \citep{Kurosawa_etal:2006,Podio_etal:2006}. There is evidence in favour of both theories; for instance the outflow model is supported by measurements of the angular momentum of jets \citep{Ray_etal:2007}, but not by the fact that discs are self-luminous, since an outflow which carries all of the angular momentum also removes all of the accretion energy. It is likely therefore that discs are outflow- or turbulence-dominated in different zones, at different times and in different objects (see \citealt{Combet_etal:2010} and refs.\ therein for a recent comparison of the two).

There is strong evidence that discs do contain strong, ordered, net poloidal flux. This is crucial for the outflow model \citep{Ouyed_etal:1997,Banerjee_Pudritz:2006,Banerjee_Pudritz:2007,Beckwith_etal:2008} but possibly also helpful in increasing the efficiency of the MRI, bringing the Shakura-Sunyaev $\alpha$ parameter produced in simulations up to the observationally-inferred value \citep{King_etal:2007,Pessah_etal:2007}. Moreover, there are direct measurements of the magnetic field in protostellar discs at various radii. \citet{Vlemmings_etal:2010} used methanol masers in the massive-protostar system CepheusAHW2, finding a field of $23$ mG at a disc radius of $\sim 1000$ AU and a local mass-to-flux ratio of $\lambda \sim 1.7$. The field is large-scale and has a direction similar to that in the surrounding cloud. Using meteorites, \citet{Levy_Sonett:1978} measured $3$ G at radius $1$ AU in the proto-solar system, translating to $\lambda\sim5$, using observations (e.g.\citealt{Wilner_Lay:2000}) giving surface densities of order $10^4$ g cm$^{-2}$ at this radius. Finally, in the FU Orionis system \citet{Donati_etal:2005} used the Zeeman effect to measure a vertical field component of $1$ kG with a filling factor of $20\%$ at a radius of $0.05$ AU, i.e. $10R_\odot$. With a surface density of $\sim4\times 10^{4}$ g cm$^{-2}$, this gives a mean mass-to-flux ratio $\lambda\approx 0.1$.\footnote{Note that $\beta \sim \lambda^2 (h/r) M_\ast/M_{\rm disc}$ where $\beta=8\pi P/\bar{B^2}$; in a disc therefore it is possible to have both $\beta>1$ and $\lambda<1$.} 
 This poloidal field is accompanied by a somewhat weaker toroidal component, contrasting with local dynamo models where a predominantly toroidal field is generated {\it in situ}. Although these observational results are approximate and relate to different radii in different objects, we see that discs are close to magnetically critical even very close to the protostar -- there is no evidence for $\lambda$ anywhere near as high as in stars ($\lambda=10^3 - 10^8$, see below).

It is important to note that a small $\lambda(r)=2\pi G^{1/2}\Sigma(r)/B_z(r)$ in the disc does not necessarily mean that mass and flux are being advected inwards in that same ratio (and if the aforementioned value of $\lambda\approx0.1$ is true, this is impossible). The mass might be slipping inwards past the magnetic field lines; all we can say with certainty is that the flux present in the disc was advected inwards at some point in the past. In the turbulent-disc scenario, \citet{vanBallegooijen:1989} used geometrical arguments to show that the turbulent diffusion should cause the magnetic field to diffuse outwards relative to the gas at the roughly same speed at which the gas moves inwards. There may be ways around this: {\mk for instance,} \citet{Spruit_Uzdensky:2005} proposed a mechanism to advect flux inwards in discrete clumps, {\mk which is} perhaps supported by the aforementioned observations of \citet{Donati_etal:2005}. 
 In contrast, in the disc-wind/jet model, one expects {\it prima facie} that flux is advected inwards since accretion is fast and there is no turbulent diffusion.

If in the steady state the star is accreting mass and flux in the ratio $\lambda_\ast$, then mass and flux must be passing through each surface of constant radius in the disc in the same ratio (ignoring outflow), even though the local ratio will in general be much lower $\lambda(r)\ll\lambda_\ast$, requiring almost perfect slippage at all radii. Since the properties of the disc vary significantly over the large range in radii, it is natural to infer that there is some feedback mechanism which prevents flux from moving inwards faster than it can be absorbed into the star, which in turn will limit the mass accretion rate to that at which the mass can effectively slip past the field lines. This mechanism could work locally, `feeling' some quantity such as the radial gradient in field strength, or it could be cyclic. Since observations show that $\lambda(10R_\odot)\ll \lambda_\ast$ it seems likely that this feedback originates from a bottleneck in the central region where it is noteworthy that a high ionisation fraction renders ambipolar and Ohmic diffusion ineffective. Finally, note that we do not know the value of $\lambda_\ast$ while an embedded star is accreting --  it may well be very much lower than the mass-to-flux ratio of stars which have finished accreting. It is possible that much of the flux is lost once the main accretion phase is over and the star has become detached from its surroundings.

The purpose of this paper is to show that any {\it excess flux} can be destroyed once the protostar forms. {\mk Excess flux is taken to mean the difference between the flux accreted onto the protostar and the flux later observed on the star}. The easiest way to imagine how this can happen is to assume first that all flux which survives diffusive processes during the supercritical core collapse is accreted onto the {\mk protostar, which therefore at least initially contains the entire accreted flux. The star then} becomes magnetically `detached' from its parent cloud, and then after accretion has finished the flux is destroyed according to the mechanisms described in section \ref{sec:destruction}. One might actually expect these mechanisms to work continuously as accretion is ongoing, so that the flux of the infalling material is lost as soon as it arrives at the protostar; this scenario is discussed in section \ref{sec:ongoing}.

\subsection{Conditions in the proto- and main-sequence star}

Of crucial importance for the following discussion is the appearance of convection, which appears in protostars when a temperature of $10^6$ K is reached and deuterium fusion begins. If the mass remains below $0.4M_\odot$, the star remains on the fully-convective Hayashi track all the way onto the main-sequence (MS) when hydrogen ignites. Above $0.4M_\odot$ a radiative interior eventually develops, either after accretion has ceased, whereupon the star enters the Henyey track \citep{Henyey:1955,Palla_Stahler:1993} and moves to the left on the HR diagram, or in a more massive protostar while accretion is still ongoing -- {\mk deuterium burns in the convective shell at the same rate at which it is accreted. Above about $4M_\odot$ convection} retreats right to the surface of the star \citep{Palla_Stahler:1993} and may dissappear entirely or have little effect on the accreting gas. This marks a fundamental difference between high-mass and low-mass stars, with the all the material in stars below $4M_\odot$ having experienced convection, but only the inner $4M_\odot$ of more massive stars having experienced convection. In addition, it is possible that very massive protostars somehow bypass the fully convective phase during growth up to $2M_\odot$, retaining a radiative interior at all times.

On the MS, stars below $0.4M_\odot$, which are fully convective, and stars between $0.4$ and $1.5M_\odot$, which have a convective envelope, display fluctuating dynamo fields. It is not obvious from first principles whether a magnetic-convective steady state should depend on the initial conditions; all we can say is that no such dependence is apparent from the observations, which do however show a clear positive correlation between rotation speed and magnetic activity (\citealt{Pizzolato_etal:2003}; see also \citealt{Morin_etal:2010}).

In contrast, in MS stars with a radiative envelope ($>1.5M_\odot$) the observations of magnetic fields are the following. Of the less massive stars (up to $6M_\odot$) some fraction ($\sim 10\%$, see e.g.\ \citealt{Power:2007}) display a weak, large-scale field of $0.2-30$ kG; the so-called Ap and Bp stars. At higher masses (O and early B) it seems that a similar fraction is magnetic \citep{Grunhut:2011}. The fields strengths measured correspond to a {\mk mass-to-flux ratio of $\lambda\lesssim10^{-3}$ }assuming similar interior and surface field strengths. These fields are presumably in equilibrium, since there is no possibility of a contemporary dynamo. The rest of the {\mk population, however, seem to have fields at the gauss-level or lower \citep{Lignieres:2009,Petit_etal:2011}, which may be for} instance produced by a subsurface-convective-dynamo mechanism (Cantiello \& Braithwaite, in prep.) or be evolving decaying dynamically on a timescale given by $\tau_{\rm A}^2/P$ where $\tau_{\rm A}$ and $P$ are the Alfv\'en timescale and rotation period (Braithwaite \& Cantiello, in prep.) In any case, the important point is that in contrast to convective stars, the magnetic properties of radiative stars do depend on the initial conditions. For a recent review of magnetism in main-sequence stars see \citealt{Donati:2009a}. Fields can also be observed directly on stars which are still accreting; in both convective and non-convective stars the fields may not be very different from those on the main-sequence (see e.g.\ \citealt{AlecianE:2009}).

This agrees with what one expects theoretically. {\mk In a convective star, the magnetic field reaches a steady state independent of the initial conditions. In a radiative star the magnetic field evolves on a dynamic timescale until it finds some stable equilibrium, the geometry and strength of which does depend} on the initial conditions.

\section{Destruction of flux in a star}\label{sec:destruction}

\begin{figure}
\includegraphics[width=1.0\hsize,angle=0]{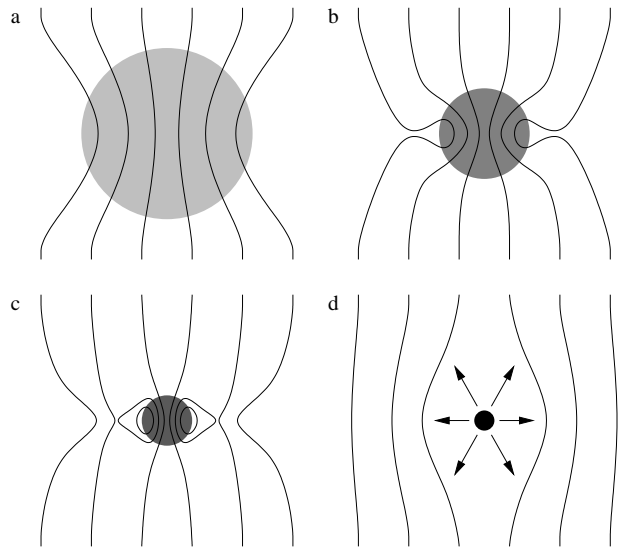}
\caption{The magnetic disconnection of a collapsing core/protostar from the rest of the cloud (for clarity, very simplistic and not to scale). (a) A collapsing overdense region draws field lines together, the classical `hourglass' picture. (b) Further collapse. (c) Reconnection occurs outside the protostar, which becomes disconnected from the cloud. (d) A super-Alfv\'enic wind ensures that the star is completely causally disconnected from its surroundings, and the flux in the star can now be destroyed. The wind expels the surrounding cloud material.}
\label{fig:hourglass}
\end{figure}

Once accretion has stopped and the star has developed a super-Alfv\'enic wind, the star is `detached' from the parent cloud and its flux no longer needs to be conserved. {\mk In fact, this detachment may happen earlier than the complete cessation of accretion, when the accretion rate decreases, and the star could possibly detach and reattach many times before final detachment, in response to FU Orionis accretion episodes.} A possible sequence of events is illustrated in fig.~\ref{fig:hourglass}, and although the actual path between stages (a) and (d) is uncertain, the detail is unimportant. I now look at two mechanisms to destroy flux in the idealised case of a fully detached star, which work respectively via MHD instability and reconnection to equilibrium inside a radiative star (section~\ref{sec:non-conv}) and buoyant expulsion from the interior of a convective star (section~\ref{sec:conv}). Afterwards, in section \ref{sec:ongoing}, I look at the more realistic scenario where these mechanisms proceed while the star is still accreting.

\subsection{Non-convective stars}\label{sec:non-conv}

\begin{figure}
\hspace{1.3cm}\includegraphics[width=0.68\hsize,angle=0]{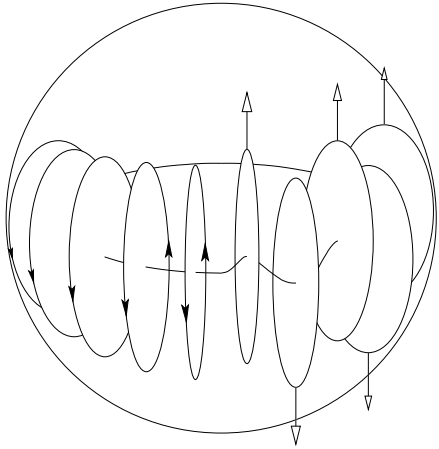}
\caption{Form of the instability in a poloidal field. On the left is shown the equilibrium field-line loops; on the right the form of the unstable motions at some particular azimuthal wavenumber.}
\label{fig:fr}
\end{figure}

In the simplest picture, a stably-stratified star initially contains a symmetric, dipole-like purely-poloidal field which was inherited from the cloud. Any purely poloidal field is subject to an MHD instability \citep{Markey:1973,Markey:1974,Flo_Rud:1977,Braithwaite:2007,Marchant:2010} which is similar to the instability in a pair of initially aligned bar magnets which rotate until they lie with opposite poles next to each other. The free energy is the volume integral of $B^2/8\pi$ outside the magnets, which drops as a result of the realignment. A fluid star can be thought of as a collection of bar magnets; the magnets do not stay aligned and the instability grows on the dynamical (Alfv\'en) timescale, around $10$ years in a solar-type star with a field strength of $1$ kG -- in any case much shorter than any formation timescale. See fig \ref{fig:fr}. Eventually length scales become small enough so that magnetic flux is destroyed inside the star, resulting in the complete destruction of the magnetic field. Note that even if the field is not purely poloidal but is `wound-up' to some degree, there is nothing to stop it unwinding first and then developing this instability.

If the field is not symmetrical as above but has some net `twist' then MHD instability will not destroy it entirely. Instead, it will evolve on the Alfv\'en timescale into some stable equilibrium. The main factors which determine the nature of the resulting equilibrium are a) the net twist, or more formally the magnetic helicity and b) the initial radial distribution of magnetic flux.

\subsubsection{Helicity}\label{sec:helicity}

Let us imagine the relaxation to magnetohydrodynamic equilibrium in a star of radius $R_\ast$ which contains a magnetic field of energy $E=(4\pi/3)R^3 B^2/8\pi$, where $B$ is the r.m.s.\ magnetic field. In the following, quantities initially and at (finally) equilibrium are marked with the subscripts i and f respectively. We can say the following about the equilibrium state.

The reconnection destroys magnetic energy on small length scales but has little effect on the magnetic helicity, a global quantity defined as the volume integral of the scalar product of the magnetic field with its vector potential $H\equiv(1/8\pi)\int\!{\bf A}\!\cdot\!{\bf B}\,{\rm d}V$. It can be shown that in the case of infinite conductivity, helicity is conserved \citep{Woltjer:1958}. Helicity has units of energy $\times$ length and so is present more in the larger structures than is the energy -- and it is {\it approximately} conserved during reconnection taking place on small scales, a property which has been very useful in many contexts from the laboratory \citep{Chui_Moffat:1995,Hsu_Bellan:2002} to the solar corona \citep{Zhang_Low:2003}. Therefore:
\begin{equation}\label{eq:hel_cons}
H_{\rm f} \approx H_{\rm i}.
\end{equation}
Consideration of dimensions gives us
\begin{equation}\label{eq:hel_val}
|H_{\rm f}| = \psi_{\rm f} R_\ast E_{\rm f}, 
\end{equation}
where $\psi_{\rm f}$ is a dimensionless parameter of the equilibrium; 
we can solve for $E_{\rm f}$ and $H_{\rm f}$ once we know its value. In a large-scale equilibrium $\psi_{\rm f}$ should be of order unity; in a small-scale equilibrium it will be smaller but note that during a relaxation to equilibrium at constant helicity an equilibrium with higher $\psi_{\rm f}$ is always favoured since it represents a lower energy state.

So, the energy and strength of the magnetic field on the main sequence is determined by the helicity of the parent cloud, which can be expressed as $|H_{\rm cl}|=\psi_{\rm cl}R_{\rm cl}E_{\rm cl}$ where $\psi_{\rm cl}$ is the degree of twist in the cloud, which can have any value from $0$ to order plus or minus unity (helicity is either positive or negative depending on the sense of the twist). This logic was recently confirmed numerically in the context of intergalactic bubbles \citep{Braithwaite:2010}, where a relaxation to equilibrium was found approximately to conserve helicity, upon which the final equilibrium energy therefore depends.

Finally, note that in the standard hourglass model, helicity is zero and so one expects the field to disappear completely after disconnection from the cloud. A non-zero helicity requires some asymmetry between the hemispheres.

\subsubsection{Stratification}

\begin{figure}
\includegraphics[width=1.0\hsize,angle=0]{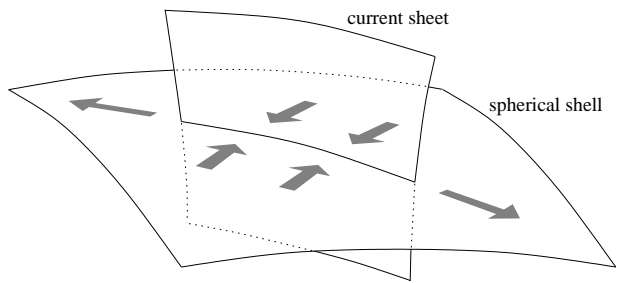}
\caption{Reconnection on a spherical shell where a discontinuity appears in the magnetic field. The grey arrows represent the velocity field; note that the motions are essentially two-dimensional.}
\label{fig:spherical_shell}
\end{figure}

Assuming that the magnetic energy is much less than the gravitational, a positive radial entropy gradient prevents the gas from moving significantly in the radial direction (we speak of a stably-stratified star), so that during the relaxation to equilibrium the gas motion is confined to spherical shells.\footnote{In addition, the motion is approximately incompressible, so the velocity field has just one degree of freedom.} This means that the absolute flux through a spherical shell cannot increase during relaxation to equilibrium. This can be seen by considering two or more regions on a spherical shell of positive and negative radial component of magnetic field $B_r$; see fig.\ \ref{fig:spherical_shell}. As the fluid moves around on the spherical shell discontinuities (current sheets) develop between regions of different $B_r$ (not just between positive and negative $B_r$ but between any differing values) and the result of these sheets is that gas from either side with initially different $B_r$ ends up with some $B_r$ which is an average of the two. Indeed in general, dissipative processes can only lead to a {\it drop} in the spherical surface integral $\oint |B_r| \rd S$ at any given radius. Therefore, if the star initially contains a strong magnetic field in the centre and a weak field at the surface, which is what one might expect if each fluid element conserves its flux during accretion and compression so that $B\propto\rho^{2/3}$, then the resulting equilibrium will also have only a weak field at the surface. It is possible for a star to `hide' a large magnetic energy in its interior but show nothing at the surface. In fact, if $B\propto\rho^{2/3}$ thoughout the star then Ap stars, which have magnetic to thermal energy density ratios at the surface of between $1$ and $100$ (i.e. the plasma-$\beta$ is $0.01$ to $1$), have global $E_{\rm mag}/|E_{\rm grav}|$ ratios of the about same value. Of course, a global ratio of unity or greater is impossible since the star would be gravitationally unbound. 
However, the lower conductivity near the surface means that a star which initially has $B\propto\rho^{2/3}$ will undergo diffusion which leads to the field in a layer below the surface relaxing to a potential field, meaning that the field strength at the surface will be similar to that at the bottom of the layer, where the plasma-$\beta$ is much greater than unity; the $B\propto\rho^{2/3}$ relation continues to hold deeper in the interior. For instance, in a $2M_\odot$ ZAMS star the potential-field layer should grow to a depth of $\sim0.05R_\ast$ after a time $10^7$ yr (calculated by equating $t$ to $L^2/\eta$ where $L$ is depth and $\eta$ the magnetic diffusivity at that depth); at that depth the density is $\sim10^4$ times its photospheric value; if a field strength of $3$ kG seen on the surface is roughly constant down to that depth and then $B\propto\rho^{2/3}$ below that, the global ratio $E_{\rm mag}/|E_{\rm grav}|\sim5\times10^{-4}$. Most of the flux would be hidden within the star. 
 Alternatively the field strength in the interior may depend much less steeply on density; in fact if one imagines building a (radiative) star by accretion, adding spherical shells from inside to outside (as seems likely from entropy considerations), it is difficult to see how the fluid elements could possibly retain their original flux without a very unlikely and unstable magnetic field configuration. In any case, there is still potential for the field strength deep in the interior to be much greater than that seen on the surface. 
 
\subsection{Convective stars}\label{sec:conv}

Alternatively, if at some time the star becomes convective then the magnetic field will tend to rise to the surface, its energy being dissipated by reconnection in the atmosphere. While the magnetic energy density is greater than the convective, as might be expected in a protostar at least initially, the important condition is not the convective {\it motion}, but the near-isentropic state which this motion maintains. {\mk Deviations from uniform entropy are also unimportant when the field is above equipartition with the convective kinetic energy -- buoyancy effects are dominated by the magnetic field.}

\subsubsection{Buoyant loss of magnetic flux}

A magnetic field in an isentropic star has a tendency to rise towards the surface \citep[][and refs.\ therein]{Reisenegger:2009}. Any magnetised region must be in pressure equilibrium with its surroundings, which means that its gas pressure must be lower than that in the surroundings. Since the entropy is constant we have $\rho=\rho(P)$, so that the region must also have lower density. There is no way to stop buoyant rise to the surface unless reconnection can occur fast enough that a global magnetic equilibrium is reached where buoyant regions are somehow held down by magnetic tension. It is therefore informative to estimate the relevant timescales.

Consider a body of gas sitting in hydrostatic equilibrium in a gravitational potential well. The gas contains a non-equilibrium magnetic field of r.m.s. magnitude $B$ with a characteristic length scale $l$, which is initially much smaller than the size of the system $R$ (below, the case with initial $l\sim R$ is looked at). Magnetic reconnection occurs at a speed
\begin{equation}\label{eq:v_rec}
v_{\rm rec} = \alpha v_{\rm A}
\end{equation}
where the reconnection speed parameter $\alpha$ is of order unity (in the solar corona context a value $0.1$ is often assumed) and $v_{\rm A}$ is the Alfv\'en speed. As reconnection proceeds on a timescale of $l/(\alpha v_{\rm A})$, in the absence of other effects the length scale $l$ increases until some global equilibrium is reached with $l$ of the order the size of the system. However, while this reconnection is proceeding, there are buoyant motions due to magnetically-induced density variations, which also have a length scale $l$: imagine the buoyant motion of a parcel of gas of size $l$ and density $\rho$ through its surroundings of density $\rho_0$. 
 Matching the buoyancy force to the aerodynamic drag, we have
\begin{equation}\label{eq:FeF}
\rho_0 l^2 v_{\rm t}^2 \sim g|\rho_0-\rho|l^3
\end{equation}
where 
 $v_{\rm t}$ is the terminal velocity (upwards or downwards depending on the sign of $\rho_0-\rho$) and $g$ is the gravitational acceleration. It can easily be shown that the terminal velocity is achieved after the region has risen a distance $l$. 
 From the equation of hydrostatic equilibrium
 we can reorganise (\ref{eq:FeF}) to give
\begin{equation}\label{eq:vt1}
v_{\rm t} \sim \left(\frac{l}{H_p}\right)^{1/2}\,\left(\frac{|\rho_0-\rho|}{\rho_0}\right)^{1/2}\,c_0,
\end{equation}
where 
 $H_p$ is the pressure scale height and
 $c_0$ is the sound speed in the external medium (recall that $c_0^2\sim P_0/\rho_0$).
 We can assume $\rho\approx\rho_0$, so that the motion is very subsonic\footnote{In the case of X-ray cavities in galaxy clusters, we have $\rho\ll\rho_0$ and $l\approx H_p$, so 
 that $v_{\rm t}\sim c_0$. From the apparent absence of shocks in these systems we infer that the motion is in fact mildly subsonic and cannot therefore solve the cooling-flow problem with shock heating.}. Pressure balance gives\footnote{It is appropriate here to use an `isotropic' magnetic pressure equal to one third of the energy density, as with any relativistic fluid \citep[see e.g.][]{Braithwaite:2010}.}
\beq\label{eq:pmag}
P_0+\frac{B_0^2}{24\pi}=P+\frac{B^2}{24\pi}.
\eeq
This ignores the anisotropic nature of magnetic pressure; in reality neighbouring blobs will push and pull on each other in some complex way, but we limit ourselves here to looking just at the general tendency for buoyant rise. In an isentropic star we have $P=K\rho^\gamma$ so
\begin{equation}\label{eq:dpdrho}
\frac{P_0-P}{P_0}\approx\gamma\frac{\rho_0-\rho}{\rho_0},
\end{equation}
in contrast to a star with an extra thermodynamic degree of freedom where buoyant balance $\rho=\rho_0$ can be achieved despite the pressure difference: for instance in a star with an ideal-gas EOS the temperature can be adjusted. Using (\ref{eq:pmag}) and (\ref{eq:dpdrho}), (\ref{eq:vt1}) becomes (dropping factors of order unity and assuming $|B-B_0|\sim B$) 
\begin{equation}\label{eq:v_t_A}
v_{\rm t} \sim \left(\frac{l}{H_p}\right)^{1/2}\,v_{\rm A}.
\end{equation}
This corresponds to the result of \citet{Parker:75} for the rise of a flux tube in a convective zone where the hydrodynamic force associated with the convective motion can be ignored.
Comparing this to (\ref{eq:v_rec}) we see that the condition for reconnection to proceed faster than buoyant motion is
\begin{equation}
\frac{l}{H_p} \lesssim \alpha^2.
\label{eq:l}\end{equation}
It seems therefore that the relaxation of an arbitrary magnetic field with small initial $l$ to equilibrium as seen in simulations of a stably stratified star cannot occur in a star with $\rho=\rho(P)$ beyond the point where the length scale is some fraction of the scale height. If on the other hand the initial conditions have a larger length scale $l$ than that in (\ref{eq:l}), the buoyant motion is always faster than the reconnection speed and so it is impossible for the magnetic field to reorganise itself. Regions of strong field will rise to the surface while regions of weak field will sink to the centre; at or above the surface of the star (in the low-plasma-$\beta$, force-free regime) there could be reconnection and loss of magnetic helicity, eventually resulting in complete loss of the magnetic field of the star via a continuous process of magnetic buoyant convection. In principle it seems possible that in the absence of heat-driven convection, some field might be retained and form a global equilibrium\footnote{The range of available equilibria in isentropic stars, or alternatively in stars with a barotropic EOS $\rho=\rho(P)$, is more restricted than in non-barotropic stars since there is only one degree of freedom in adjusting the thermodynamic state to balance the Lorentz force, but their existence cannot be ruled out.}; however, an equilibrium not only requires non-zero helicity as in the case of stable stratification explored in section \ref{sec:non-conv}, but should eventually be destroyed by the convective motions.

However, how can one justify ignoring the convective motion in the above analysis? To suppress convection, a field must not only be coherent on scales larger than the convective cells ($\sim H_p$) but also at least at equipartition with the thermal energy \citep{Mestel:1970}, which we can assume is not the case for Hayashi-track stars. However, while the energy in the field is greater than the kinetic energy of the convection the convection is sub-Alfv\'enic and therefore makes little difference to the magnetic buoyant motions explored above.

In any case, once the magnetic field has decayed to equipartition with the convection, the convective motion can no longer be ignored, the analysis above becoming invalid. The domain of the convective dynamo has been entered, and some steady-state is presumably reached with the magnetic energy at most comparable to the convection energy. Any `memory' of the original (large) magnetic flux is presumably erased, and the main-sequence magnetic field does not depend on the conditions in the parent cloud.

\subsubsection{Numerical model of an isentropic star}\label{sec:numerical}

\begin{figure*}
\includegraphics[width=0.33\hsize,angle=0]{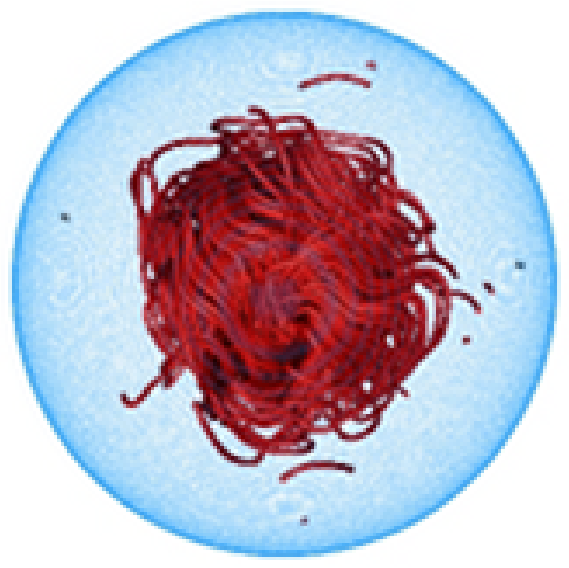}
\includegraphics[width=0.33\hsize,angle=0]{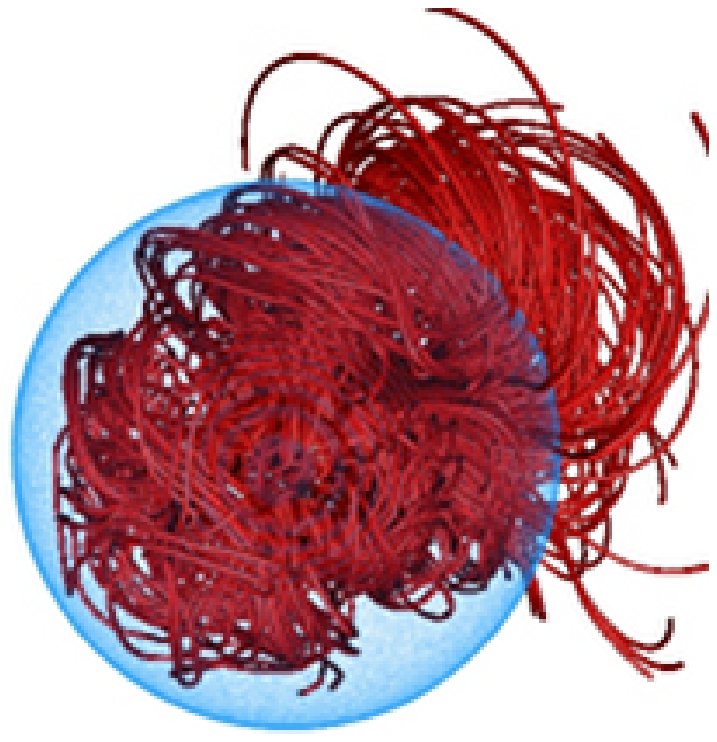}
\includegraphics[width=0.33\hsize,angle=0]{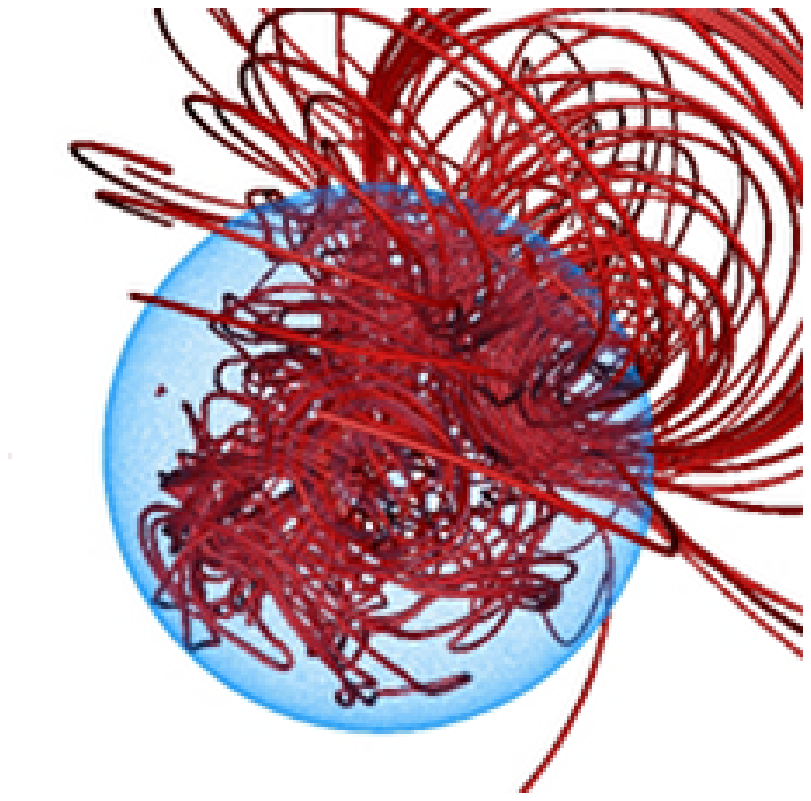}
\caption{The evolution of an initially turbulent magnetic field in a non-stably stratified star. {\mk The three snapshots are at $t/\tau_{\rm A}=0$, $3.1$ and $7.8$ where $\tau_{\rm A}$ is the Alfv\'en timescale. After some Alfv\'en times, the} field has bubbled up to the surface and shows no signs of reaching an equilbrium. The blue shading represents the stellar surface.}
\label{fig:vapor}
\end{figure*}

I now describe a numerical model of flux loss from an isentropic star. 
 The stellar model and computational setup are similar to those described in \citealt{Bra_Nor:2006}, to where the reader is referred for a fuller account of the setup of the model; a brief outline is given here. The code used is the {\sc stagger code} (\citealt{Nor_Gal:1995}, \citealt{Gud_Nor:2005}), a high-order finite-difference Cartesian MHD code which uses a `hyper-diffusion' scheme, a system whereby diffusivities are scaled with the length scales present so that badly resolved structure near the Nyquist spatial frequency is damped whilst preserving well-resolved structure on longer length scales. This, and the high-order spatial interpolation and derivatives (sixth order) and time-stepping (third order) increase efficiency by giving a low effective diffusivity at modest resolution ($192^3$ here). The code includes Ohmic and well as thermal and kinetic diffusion. Using Cartesian coordinates avoids problems with singularities and simplifies the boundary conditions: periodic boundaries are used here.

The simulations model the star as a self-gravitating ball of ideal gas ($\gamma=5/3$) of radius $R$ in hydrostatic equilibrium with radial density and pressure profiles obeying the polytrope relation $P \propto \rho^{1+1/n}$, with the index $n$ set to $3/2$ here to give constant entropy, as opposed to the radiative-star approximation $n=3$ in \citealt{Bra_Nor:2006}. Surrounding the star is a hot, poorly-conducting atmosphere, which has the effects both of {\mk forcing the magnetic field to relax into a potential (curl-free) state while keeping the Alfv\'en speed outside the star numerically convenient, since the density is not too low; in other words it is a} simple way of modelling a vacuum. Since the field strength drops by a large factor during the process being modelled, a field-upping routine is employed, which artificially amplifies the field to keep the magnetic energy constant (see \citealt{Bra_Nor:2006} for details), which greatly reduces the computational demand and, perhaps more importantly, keeps the Alfv\'en timescale much shorter than the diffusive (Ohmic) timescale. At the beginning, the star is given an initially random magnetic field containing energy at all length scales down to a limit of a few grid-spacings, and the MHD equations are integrated in time to follow the evolution of the field.

\begin{figure}
\includegraphics[width=1.0\hsize,angle=0]{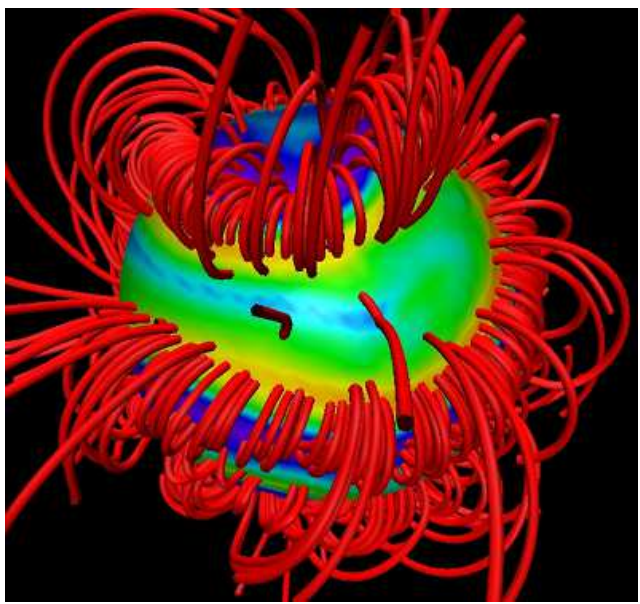}
\caption{When the field reaches the surface of the star, familar structures become visible. Here, field lines are plotted and the surface of the star is coloured according to the value of the radial field component (blue \& violet are negative, green \& yellow/orange are positive). This run is similar to the one shown in fig.\ \ref{fig:vapor}, but with a different random initialisation; this snapshot is taken after 12.2 Alfv\'en timescales have passed (corresponding to the last frame of fig.\ \ref{fig:vapor}. Flux tubes lie along the surface in apparently random patterns.}
\label{fig:vapor2}
\end{figure}

{\mk In fig.\ \ref{fig:vapor} we can see an example of the evolution of a field in this numerical model. At the beginning the field is somewhat concentrated into the interior of the star, and rises towards the surface on an Alfv\'en timescale.
Another example of this, starting with a different randomisation of the initial field, is shown in fig.\ \ref{fig:vapor2}.}

\begin{figure*}
\includegraphics[width=0.33\hsize,angle=0]{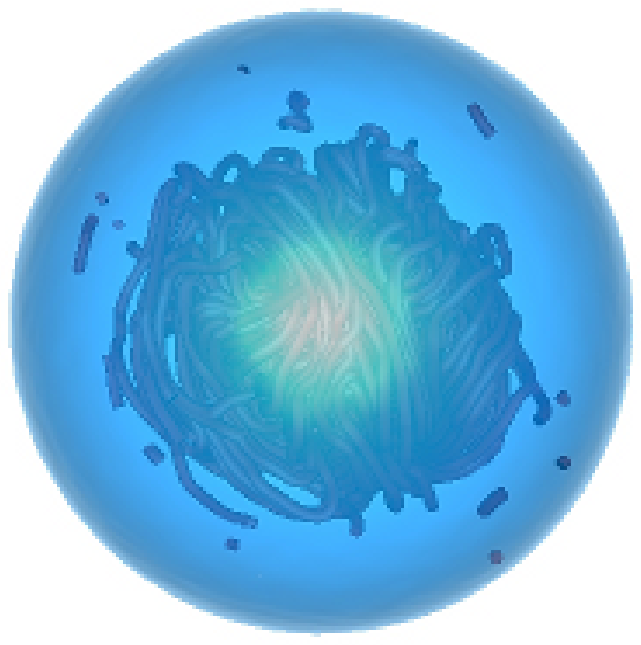}
\includegraphics[width=0.33\hsize,angle=0]{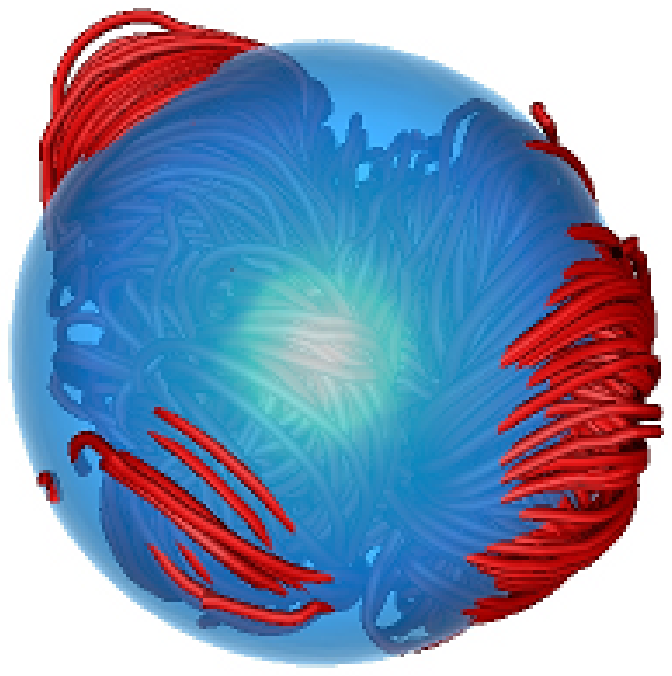}
\includegraphics[width=0.33\hsize,angle=0]{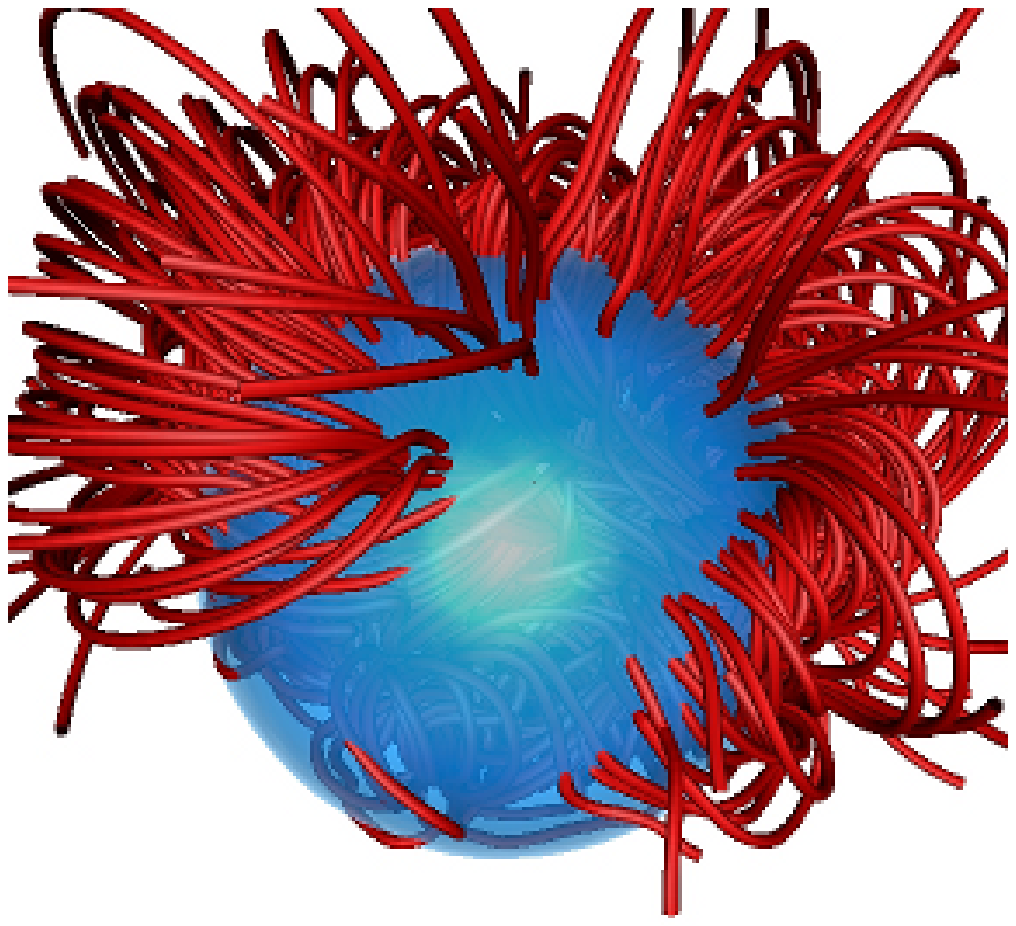}
\includegraphics[width=0.33\hsize,angle=0]{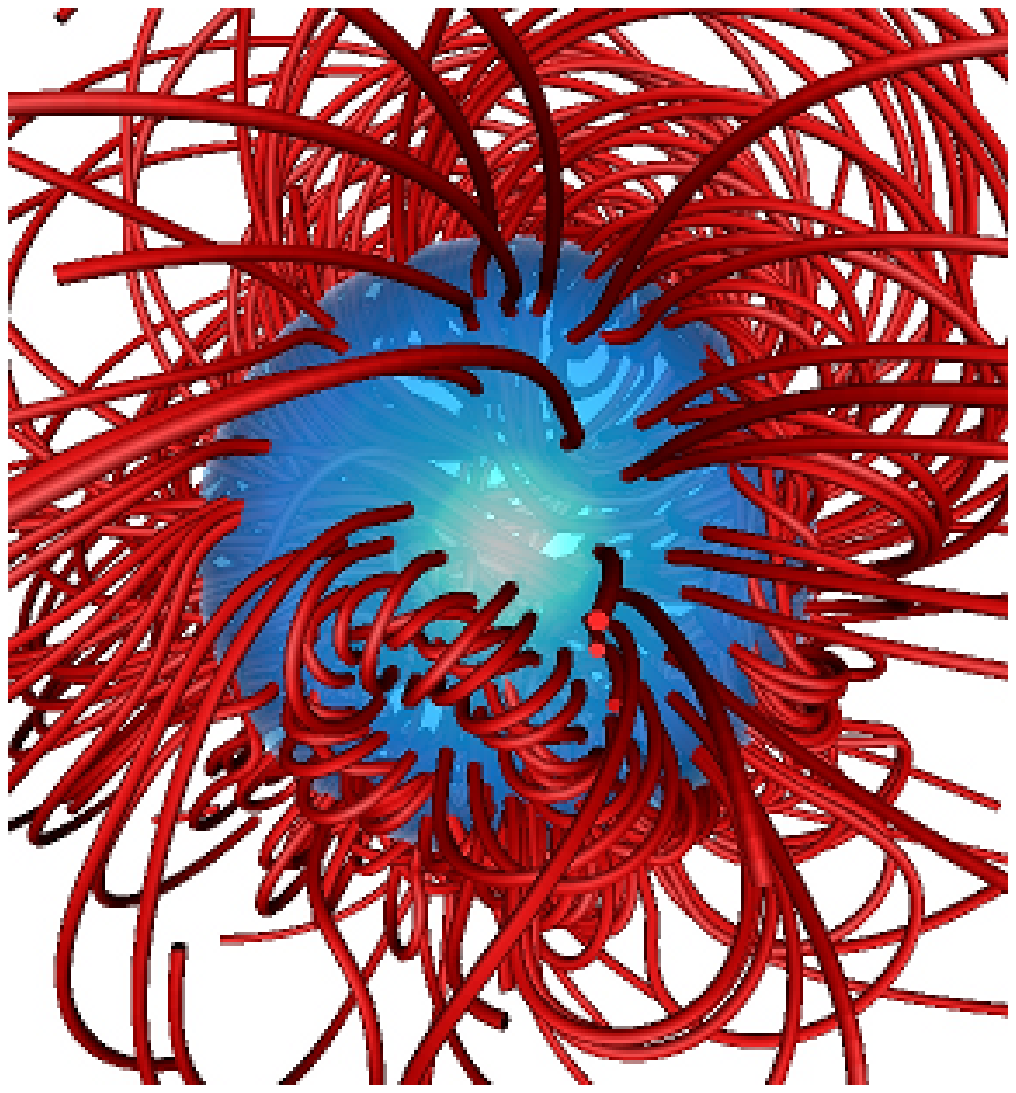}
\includegraphics[width=0.33\hsize,angle=0]{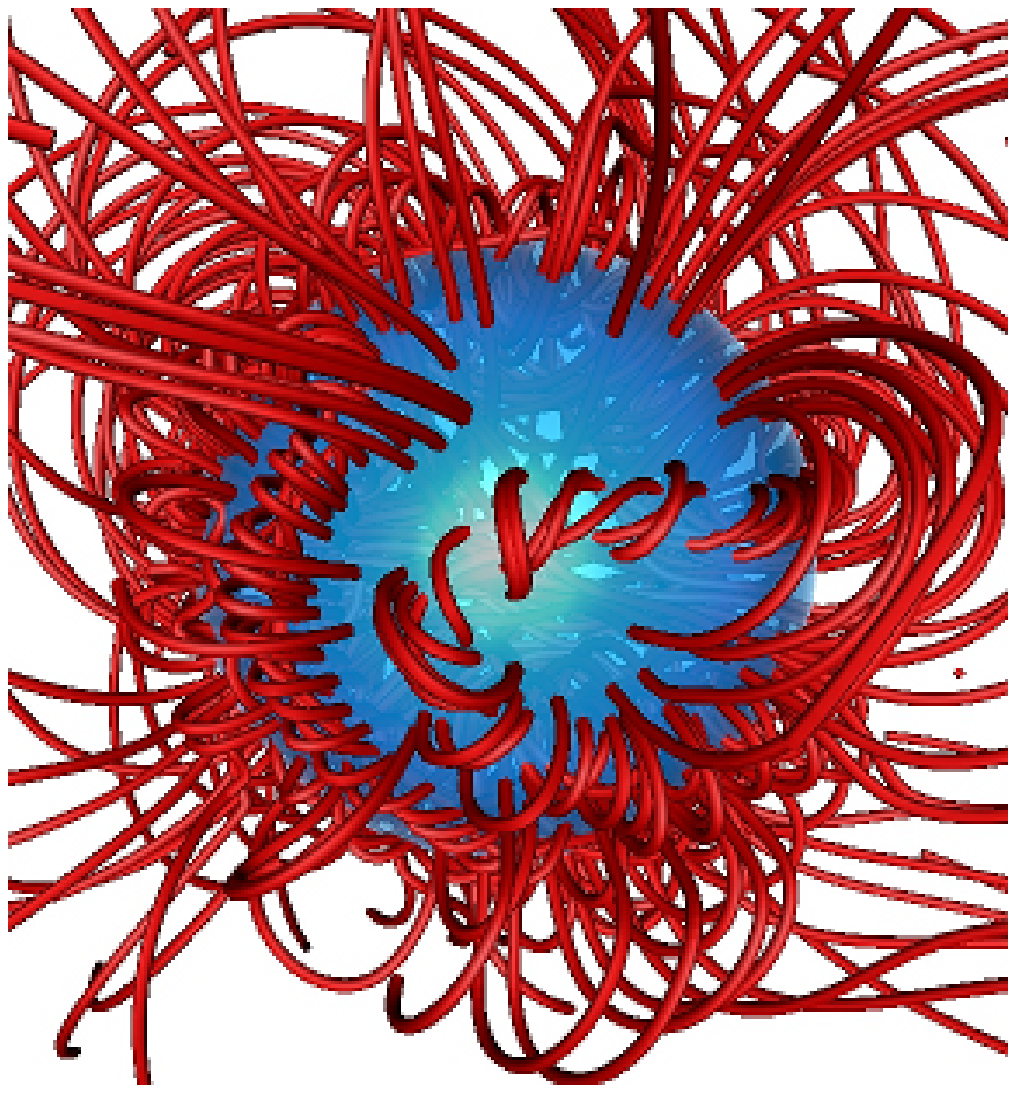}
\includegraphics[width=0.33\hsize,angle=0]{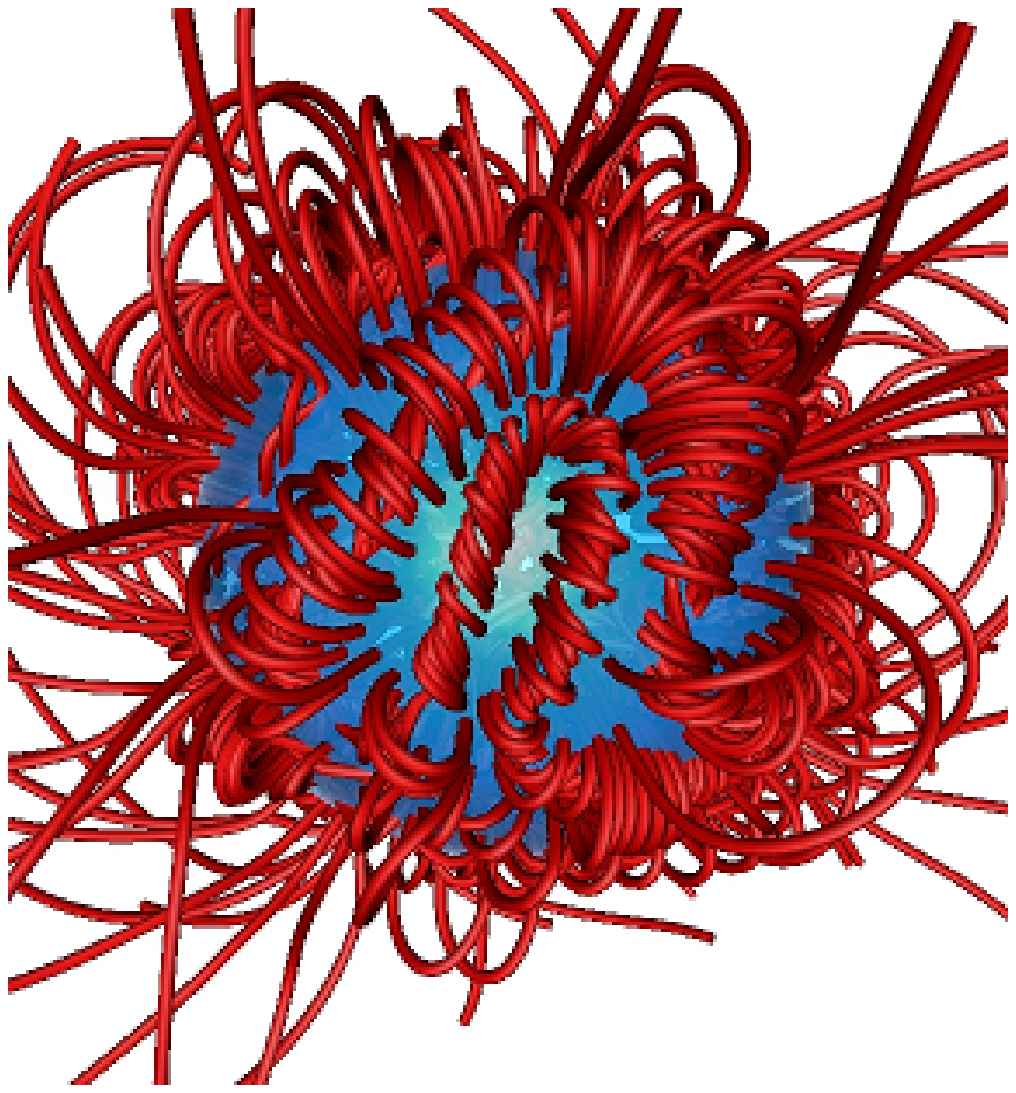}
\caption{A time sequence at $t/\tau_{\rm A}=0$, $1.6$, $4.2$, $9.4$, $13.1$ and $19.3$ (top row left to right and then bottom row left to right). The field quickly reaches the surface, but only in some places, since there must be a downwards flow in other places. Discrete flux tubes become recognisable, as in the fourth frame. In the last frame, the entire field can be thought of as discrete flux tubes. As these flux tubes are pushed upwards by buoyancy they lose their axial component into the atmosphere which causes them to become longer and thinner. This is similar to process in stably stratified stars, except that it happens on a dynamic timescale rather than a diffusive timescale.}
\label{fig:vapor3}
\end{figure*}

{\mk Taking a closer look, at some point in time some recognisable structure becomes visible, in the form of flux tubes lying quasi-horizontally around the surface of the star. These resemble the structure of non-axisymmetric equilibria found in previous simulations \citep{Braithwaite:2008} when a similarly random, but less centrally-concentrated initial field (i.e. with significant flux threading the surface at $t=0$) is left to evolve in a {\it stably-stratified star} -- horizontally-lying flux tubes arranged in some meandering pattern below the surface of the star\footnote{Note that in the simulations it is difficult to define the precise location of the stellar surface, as the transition from isentropic and high electrical conductivity to isothermal and low electrical conductivity takes place over several grid spacings}; these tubes evolve on a timescale given by the Ohmic timescale and/or a large multiple of the thermal timescale (whichever is shorter) under the stellar surface, which can of course be significant compared to, or longer than, a main-sequence lifetime. In contrast, the similar structures found in the current study are evolving on a shorter timescale related to the Alfv\'en timescale in the stellar interior, due to the buoyant force pushing them upwards into the atmosphere of the star, and the Ohmic timescale {\it above} the stellar surface, as the tubes decay due to (relatively low) finite conductivity. 
 It is not immediately obvious what this timescale should be, except that it will be rather small compared to the evolution timescale of a protostar. In the simulations, the exterior has a much lower conductivity than the interior, which forces the exterior field to relax to a potential (curl-free) configuration, but it is impossible to make this effect quantitatively realistic in the numerical model while dynamic-timescale processes are ongoing. In all the simulations the decay timescale is a few Alfv\'en timescales. In reality the decay above the surface (in the low plasma-$\beta$ regime) should (helped by convective motions absent from these simulations) proceed on a dynamic timescale in localised reconnection zones, as it does in the solar corona.

In fig.\ \ref{fig:vapor3} we can see the process in a little more detail -- in a time animation from a third run with a different random initialisation. Several runs were done, with different random initialisations and with different initial radial profiles of magnetic energy, but in each run the result was essentially the same.}

{\mk Broadly speaking, these numerical results confirm} the prediction made in section \ref{sec:conv}. This behaviour is distinct from that of the same magnetic field in a stably-stratified star, where there is no transport of material in the radial direction \citep{Bra_Spr:2004,Braithwaite:2008} and consequently no transport of radial flux from the interior to the surface. {\mk It seems therefore that an MHD equilibrium cannot be reached from these initial conditions in an isentropic star; however it is} impossible to rule out at this stage that different initial conditions, perhaps with large length scales and a large magnetic helicity might lead to an equilibrium. {\mk The range of equilibria in an isentropic star is however more restricted than that available in a stably-stratified star, and the possibility that an equilibrium could be reached from any realistic initial conditions seems rather unlikely.} In addition, the convective motions driven by heat flux, which are not included in these simulations, would presumably prevent equilibrium formation. These questions will be studied more fully in a forthcoming publication.

\subsection{Ongoing flux loss during accretion}\label{sec:ongoing}

We have seen how flux can be destroyed once the protostar has stopped accreting and is fully detached from its surroundings, but it seems plausible that something similar to these two processes remove flux in accreted material as, or soon after, it is accreted. This is possible because unless the field is already extremely weak the Alfv\'en timescale on which these mechanisms work is much shorter than the accretion timescale. However, since we do not observe the star until the main accretion phase is over, it is difficult to distinguish observationally between ongoing flux destruction during the embedded accretion phase and flux destruction only at the end of that phase when the star becomes detached.

In the magnetospheric accretion model \citep{Koenigl:1991,Long:2008}, there is a gap (with low plasma $\beta$) between the star and the inner edge of the disc; gas is channelled from the inner edge along field lines anchored around the magnetic poles of the star. The bulk of the stellar flux self-connects within the magnetosphere, so the stellar field is essentially detached from the surroundings -- it is implicitly assumed in this model that some mechanism exists to break field lines between the star and the parent cloud. 

In addition, note that the field strength drops off sufficiently quickly with increasing distance from the star that the magnetic energy in the vicinity of the star is much greater than that further away. [More precisely, the magnetic energy per unit radius as a function of radius ($E_{\rm mag}(r)=B^2r^2/2$) must drop off faster than log normal, meaning that $\rd \ln B/\rd\ln r<-3/2$; the values for dipole and split-monopole fields are $-3$ and $-2$ respectively.] This means that energy liberated by reorganisation of the field inside and immediately outside the star outweighs any energy required to change the geometry of the field further away. 

In the general case where a net non-zero (but probably small) helicity has been accreted, a radiative star must contain an equilibrium which evolves quasistatically as a result of the addition of new mass, flux and helicity. Imagine that the star has a weak dipole field and accretes material with magnetic field aligned to that dipole: the material can simply flip over until its field is antiparallel to that of the star (similarly to the case of two bar magnets in section \ref{sec:non-conv}) whereupon its field is annihilated against that of the star; the field lines originally connecting the blob to the outside world will reconnect so as to bypass the star. In this way a lower energy state is reached, magnetic energy being converted into heat. In fact, since the flux of the accreted material is not conserved, only the accreted mass and helicity are relevant for the evolution of the star and its magnetic field.

If on the other hand the star is convective or has a significant convective envelope, its field will be in a dynamo-powered steady state. Highly magnetised material arriving at the surface will stay at the surface until both its entropy and magnetic buoyancy can be reduced by mixing  and reconnection. Again, the accreted flux is not relevant, but here also the accreted helicity is not necessarily conserved, depending on the properties of the dynamo.
 
If these mechanisms can prevent significant flux from building up in the star, the accretion disc will not be able to drag magnetic flux inwards; the magnetic field lines necessarily find some way of slipping outwards relative to the inspiralling gas. This may be facilitated by an outwards-directed Lorentz force resulting from the build up of flux towards the centre. When the disc first forms it must initially be able to drag flux inwards, accounting for the large flux observed in the inner disc; at some point that inwards drag switches off or is much reduced as magnetic pressure builds up in the centre.

\section{Discussion}\label{sec:discussion}

Let us assume for the moment that a significant fraction of the original flux survives transport into the protostar. In low-mass ($<0.4M_\odot$) protostars which have not yet become fully convective, it is possible that the flux is destroyed while accretion continues, but at the very latest when it ceases. If it is possible to measure the magnetic field on a low-mass pre-MS star after significant accretion has ended but before deuterium ignition, one should see a global MHD equilibrium rather like those seen in radiative stars on the MS. It is probably difficult or impossible to measure the magnetic field on low-mass protostars which are still accreting significantly, but this measurement could distinguish between ongoing flux destruction and destruction only after complete detachment from the surroundings. In principle flux destruction could begin as soon as the Alfv\'en timescale becomes less than the collapse timescale, i.e. when the first core forms and the collapse becomes sub-Alfv\'enic. Once deuterium ignites however, and the star becomes fully convective -- which it remains throughout the main-sequence -- the star is expected to lose all of its original flux and helicity and with it the `magnetic initial conditions' from the parent cloud. This explains why the magnetic fields of low-mass stars appear to depend only on rotation speed. Solar-type stars (eventual mass $0.4 -1.5M_\odot$) also pass through a fully-convective phase before a radiative core develops; in this envelope a dynamo similar to that in low-mass stars is expected, and observed. Inside the radiative core, one might expect to find the remnant of the previous convective dynamo in an equilibrium whose strength depends on the helicity generated by (or surviving) this dynamo. Note that for a dynamo to generate a net helicity from zero initial helicity there must be symmetry-breaking or positive feedback; alternatively an initial net helicity inherited from the molecular cloud may persist through the dynamo phase if the phase does not last too long, or may perhaps be amplified. In any case, the residual field is unlikely to be stronger than equipartition with the previous convective energy.

Intermediate-mass stars ($>1.5M_\odot$) also have a fully-convective protostellar phase 
, but convection in the envelope eventually dies away. 
 In the radiative envelope we are therefore seeing either formerly convective material, which has lost all `memory' of its initial magnetic field, as in the solar-type stars, or material which accreted onto the star after significant convection disappeared, in which case the accreted helicity may survive. Thus the difference between magnetic field strengths in intermediate-mass MS stars and the bimodality between magnetic and non-magnetic stars could be at least partially due to differences in the accretion history; the non-magnetic stars would have accreted all of their material while still convective, while the magnetic stars would have become radiative while accretion continued and were able from that point onwards to accumulate magnetic helicity from the accreted gas. This is possibly connected to the observational result that the magnetic fraction increases with mass, from $\sim1\%$ at $1.5M_\odot$ to $\sim20\%$ at $6M_\odot$ \citep{Power:2007}. In more massive stars 
 deuterium fusion contributes relatively little to the energy and convection may be unimportant, in which case we do expect to see magnetic equilibria which depend on initial conditions. However, if a star forms from a relatively symmetrical cloud then helicity should be small and we expect almost all of the magnetic energy to be lost during the relaxation to equilibrium. An equilibrium can then survive essentially unchanged for the entire main-sequence, owing to the long Ohmic timescale ($\gtrsim10^{10}$ yr), although it is possible that rotation-induced circulation has some effect over main-sequence timescales.
 
An important unsolved question therefore is on what timescale can a convective dynamo generate or destroy helicity. It is plausible that a dynamo approaches a true steady state (erasing initial conditions) on the diffusive timescale associated with the length scale of the convection (Brandenburg, priv.\ comm.) which may be rather longer than a pre-MS convective phase, but probably shorter than the ages of low-mass MS stars.

Another unsolved problem is that of the enormous range in field strengths seen in stars with radiative envelopes. It has been suggested \citep{Zinnecker:2007,Maitzen:2008,Bogomazov:2009,Ferrario:2009} that magnetic stars are merger products, the merger resulting in differential rotation and some kind of dynamo activity. This scenario would explain the lack of magnetic stars in short-period binaries \citep{Abt:1973}. However, this could also be explained by magnetic inhibition of fragmentation, as is seen in simulations \citep{Machida:2008}; also there is the issue of the lack of universal magnetic fields in blue stragglers. In any case, it seems likely that the magnetic stars are unusual in some sense and that the normal state of affairs is for the flux to drop to that corresponding to less than $1$ gauss.

\section{Conclusions}\label{sec:conc}

I have examined mechanisms to destroy flux in a protostar, with a view to 
 shedding some light on the observational fact that the magnetic flux in main-sequence stars is very small compared to that in molecular cloud cores in star-forming regions. {\mk It is uncertain how much of the original flux is already removed from the gas by various diffusive processes during collapse and accretion, and the mechanisms explored here can remove flux which survives these processes and is accreted onto the protostar.} The mechanisms can be categorised according to whether they operate in convective or non-convective protostars. In convective stars, the magnetic field rises buoyantly to the surface of the star on a dynamical timescale (an Alfv\'en timescale, $\sim10$ yr in a solar-type star with a $1$ kG field or somewhat longer in a protostar with larger radius) and its energy is destroyed by reconnection in the atmosphere, much as is observed in the solar corona, until the field strength has dropped to that which can be maintained by a convective dynamo. In a radiative star (or the radiative zone of a star), an arbitrary magnetic field evolves on the same dynamical timescale into an equilibrium, the strength of which depends not on the original flux but on the magnetic helicity. A non-zero helicity requires some asymmetry in the magnetic field of the accreted material; in the standard hourglass model the helicity is zero.

In summary 
 any excess `unwanted' flux can be destroyed once the star becomes magnetically independent from its surroundings, via buoyancy and/or MHD instability and reconnection on a dynamical timescale. This should happen either while the main accretion phase is ongoing or as it comes to an end -- in any case, by the time we can observe the star it will already have lost its original flux.

{\it Acknowledgements.} The author would like to thank Leon Mestel, \AA ke Nordlund and Wouter Vlemmings for useful discussions and assistance.


\bibliography{./Biblio}

\begin{thebibliography}{}

\bibitem[\protect\citeauthoryear{{Abt} \& {Snowden}}{{Abt} \&
  {Snowden}}{1973}]{Abt:1973}
{Abt} H.~A.,  {Snowden} M.~S.,  1973, \apjs, 25, 137

\bibitem[\protect\citeauthoryear{{Alecian}, {Catala}, {Wade}, {Bagnulo},
  {Boehm}, {Bouret}, {Donati}, {Folsom}, {Grunhut}, {Landstreet}, {Marsden},
  {Petit}, {Ramirez} \& {Silvester}}{{Alecian} et~al.}{2009}]{AlecianE:2009}
{Alecian} E.,  {Catala} C.,  {Wade} G.~A.,  {Bagnulo} S.,  {Boehm} T.,
  {Bouret} J.,  {Donati} J.,  {Folsom} C.,  {Grunhut} J.,  {Landstreet} J.~D.,
  {Marsden} S.~C.,  {Petit} P.,  {Ramirez} J.,    {Silvester} J.,  2009, in
  {C.~Neiner \& J.-P.~Zahn} ed., EAS Publications Series Vol.~39 of EAS
  Publications Series, {Magnetism in Herbig Ae/Be stars and the link to the
  Ap/Bp stars}.
pp 121--132

\bibitem[\protect\citeauthoryear{{Anderson}, {Li}, {Krasnopolsky} \&
  {Blandford}}{{Anderson} et~al.}{2003}]{Anderson_etal:2003}
{Anderson} J.~M.,  {Li} Z.,  {Krasnopolsky} R.,    {Blandford} R.~D.,  2003,
  \apjl, 590, L107

\bibitem[\protect\citeauthoryear{{Bacciotti}, {Ray}, {Mundt}, {Eisl{\"o}ffel}
  \& {Solf}}{{Bacciotti} et~al.}{2002}]{Bacciotti_etal:2002}
{Bacciotti} F.,  {Ray} T.~P.,  {Mundt} R.,  {Eisl{\"o}ffel} J.,    {Solf} J.,
  2002, \apj, 576, 222

\bibitem[\protect\citeauthoryear{{Balbus} \& {Hawley}}{{Balbus} \&
  {Hawley}}{1991}]{Balbus:1991}
{Balbus} S.~A.,  {Hawley} J.~F.,  1991, \apj, 376, 214

\bibitem[\protect\citeauthoryear{{Banerjee} \& {Pudritz}}{{Banerjee} \&
  {Pudritz}}{2006}]{Banerjee_Pudritz:2006}
{Banerjee} R.,  {Pudritz} R.~E.,  2006, \apj, 641, 949

\bibitem[\protect\citeauthoryear{{Banerjee} \& {Pudritz}}{{Banerjee} \&
  {Pudritz}}{2007}]{Banerjee_Pudritz:2007}
{Banerjee} R.,  {Pudritz} R.~E.,  2007, \apj, 660, 479

\bibitem[\protect\citeauthoryear{{Beckwith}, {Hawley} \& {Krolik}}{{Beckwith}
  et~al.}{2008}]{Beckwith_etal:2008}
{Beckwith} K.,  {Hawley} J.~F.,    {Krolik} J.~H.,  2008, \apj, 678, 1180

\bibitem[\protect\citeauthoryear{{Blandford} \& {Payne}}{{Blandford} \&
  {Payne}}{1982}]{Blandford_Payne:1982}
{Blandford} R.~D.,  {Payne} D.~G.,  1982, \mnras, 199, 883

\bibitem[\protect\citeauthoryear{{Bogomazov} \& {Tutukov}}{{Bogomazov} \&
  {Tutukov}}{2009}]{Bogomazov:2009}
{Bogomazov} A.~I.,  {Tutukov} A.~V.,  2009, Astronomy Reports, 53, 214

\bibitem[\protect\citeauthoryear{{Braithwaite}}{{Braithwaite}}{2007}]{Braithwa%
ite:2007}
{Braithwaite} J.,  2007, \aap, 469, 275

\bibitem[\protect\citeauthoryear{{Braithwaite}}{{Braithwaite}}{2008}]{Braithwa%
ite:2008}
{Braithwaite} J.,  2008, \mnras, 386, 1947

\bibitem[\protect\citeauthoryear{{Braithwaite}}{{Braithwaite}}{2010}]{Braithwa%
ite:2010}
{Braithwaite} J.,  2010, \mnras, 406, 705

\bibitem[\protect\citeauthoryear{{Braithwaite} \& {Nordlund}}{{Braithwaite} \&
  {Nordlund}}{2006}]{Bra_Nor:2006}
{Braithwaite} J.,  {Nordlund} {\AA}.,  2006, \aap, 450, 1077

\bibitem[\protect\citeauthoryear{{Braithwaite} \& {Spruit}}{{Braithwaite} \&
  {Spruit}}{2004}]{Bra_Spr:2004}
{Braithwaite} J.,  {Spruit} H.~C.,  2004, \nat, 431, 819

\bibitem[\protect\citeauthoryear{{Chui} \& {Moffatt}}{{Chui} \&
  {Moffatt}}{1995}]{Chui_Moffat:1995}
{Chui} A.~Y.~K.,  {Moffatt} H.~K.,  1995, Royal Society of London Proceedings
  Series A, 451, 609

\bibitem[\protect\citeauthoryear{{Combet}, {Ferreira} \& {Casse}}{{Combet}
  et~al.}{2010}]{Combet_etal:2010}
{Combet} C.,  {Ferreira} J.,    {Casse} F.,  2010, \aap, 519, A108+

\bibitem[\protect\citeauthoryear{{Crutcher}, {Troland}, {Lazareff}, {Paubert}
  \& {Kaz{\`e}s}}{{Crutcher} et~al.}{1999}]{Crutcher_etal:1999}
{Crutcher} R.~M.,  {Troland} T.~H.,  {Lazareff} B.,  {Paubert} G.,
  {Kaz{\`e}s} I.,  1999, \apjl, 514, L121

\bibitem[\protect\citeauthoryear{{Davidson}, {Novak}, {Matthews}, {Matthews},
  {Goldsmith}, {Chapman}, {Volgenau}, {Vaillancourt} \& {Attard}}{{Davidson}
  et~al.}{2011}]{Davidson:2011}
{Davidson} J.~A.,  {Novak} G.,  {Matthews} T.~G.,  {Matthews} B.,  {Goldsmith}
  P.~F.,  {Chapman} N.,  {Volgenau} N.~H.,  {Vaillancourt} J.~E.,    {Attard}
  M.,  2011, ArXiv e-prints

\bibitem[\protect\citeauthoryear{{Desch} \& {Mouschovias}}{{Desch} \&
  {Mouschovias}}{2001}]{Desch_Mouschovias:2001}
{Desch} S.~J.,  {Mouschovias} T.~C.,  2001, \apj, 550, 314

\bibitem[\protect\citeauthoryear{{Donati}, {Paletou}, {Bouvier} \&
  {Ferreira}}{{Donati} et~al.}{2005}]{Donati_etal:2005}
{Donati} J.,  {Paletou} F.,  {Bouvier} J.,    {Ferreira} J.,  2005, \nat, 438,
  466

\bibitem[\protect\citeauthoryear{{Donati} \& {Landstreet}}{{Donati} \&
  {Landstreet}}{2009}]{Donati:2009a}
{Donati} J.~F.,  {Landstreet} J.,  2009, ArXiv e-prints

\bibitem[\protect\citeauthoryear{{Ferrario}, {Pringle}, {Tout} \&
  {Wickramasinghe}}{{Ferrario} et~al.}{2009}]{Ferrario:2009}
{Ferrario} L.,  {Pringle} J.~E.,  {Tout} C.~A.,    {Wickramasinghe} D.~T.,
  2009, \mnras, 400, L71

\bibitem[\protect\citeauthoryear{{Flowers} \& {Ruderman}}{{Flowers} \&
  {Ruderman}}{1977}]{Flo_Rud:1977}
{Flowers} E.,  {Ruderman} M.~A.,  1977, \apj, 215, 302

\bibitem[\protect\citeauthoryear{{Galli}}{{Galli}}{2009}]{Galli:2009}
{Galli} D.,  2009, \memsai, 80, 54

\bibitem[\protect\citeauthoryear{{Gillis}, {Mestel} \& {Paris}}{{Gillis}
  et~al.}{1974}]{Gillis_etal:1974}
{Gillis} J.,  {Mestel} L.,    {Paris} R.~B.,  1974, \apss, 27, 167

\bibitem[\protect\citeauthoryear{{Gillis}, {Mestel} \& {Paris}}{{Gillis}
  et~al.}{1979}]{Gillis_etal:1979}
{Gillis} J.,  {Mestel} L.,    {Paris} R.~B.,  1979, \mnras, 187, 311

\bibitem[\protect\citeauthoryear{{Girart}, {Beltr{\'a}n}, {Zhang}, {Rao} \&
  {Estalella}}{{Girart} et~al.}{2009}]{Girart:2009}
{Girart} J.~M.,  {Beltr{\'a}n} M.~T.,  {Zhang} Q.,  {Rao} R.,    {Estalella}
  R.,  2009, Science, 324, 1408

\bibitem[\protect\citeauthoryear{{Girart}, {Rao} \& {Marrone}}{{Girart}
  et~al.}{2006}]{Girart:2006}
{Girart} J.~M.,  {Rao} R.,    {Marrone} D.~P.,  2006, Science, 313, 812

\bibitem[\protect\citeauthoryear{{Grunhut}, {Wade} \& {the MiMeS
  Collaboration}}{{Grunhut} et~al.}{2011}]{Grunhut:2011}
{Grunhut} J.~H.,  {Wade} G.~A.,    {the MiMeS Collaboration} 2011, ArXiv
  e-prints

\bibitem[\protect\citeauthoryear{{Gudiksen} \& {Nordlund}}{{Gudiksen} \&
  {Nordlund}}{2005}]{Gud_Nor:2005}
{Gudiksen} B.~V.,  {Nordlund} {\AA}.,  2005, \apj, 618, 1020

\bibitem[\protect\citeauthoryear{{Heiles} \& {Troland}}{{Heiles} \&
  {Troland}}{2005}]{Heiles_Troland:2005}
{Heiles} C.,  {Troland} T.~H.,  2005, \apj, 624, 773

\bibitem[\protect\citeauthoryear{{Henyey}, {Lelevier} \& {Lev{\'e}e}}{{Henyey}
  et~al.}{1955}]{Henyey:1955}
{Henyey} L.~G.,  {Lelevier} R.,    {Lev{\'e}e} R.~D.,  1955, \pasp, 67, 154

\bibitem[\protect\citeauthoryear{{Hsu} \& {Bellan}}{{Hsu} \&
  {Bellan}}{2002}]{Hsu_Bellan:2002}
{Hsu} S.~C.,  {Bellan} P.~M.,  2002, \mnras, 334, 257

\bibitem[\protect\citeauthoryear{{Jappsen} \& {Klessen}}{{Jappsen} \&
  {Klessen}}{2004}]{Jappsen_Klessen:2004}
{Jappsen} A.,  {Klessen} R.~S.,  2004, \aap, 423, 1

\bibitem[\protect\citeauthoryear{{Keppens}, {Casse} \& {Goedbloed}}{{Keppens}
  et~al.}{2002}]{Keppens_etal:2002}
{Keppens} R.,  {Casse} F.,    {Goedbloed} J.~P.,  2002, \apjl, 569, L121

\bibitem[\protect\citeauthoryear{{King}, {Pringle} \& {Livio}}{{King}
  et~al.}{2007}]{King_etal:2007}
{King} A.~R.,  {Pringle} J.~E.,    {Livio} M.,  2007, \mnras, 376, 1740

\bibitem[\protect\citeauthoryear{{Koenigl}}{{Koenigl}}{1991}]{Koenigl:1991}
{Koenigl} A.,  1991, \apjl, 370, L39

\bibitem[\protect\citeauthoryear{{Kurosawa}, {Harries} \&
  {Symington}}{{Kurosawa} et~al.}{2006}]{Kurosawa_etal:2006}
{Kurosawa} R.,  {Harries} T.~J.,    {Symington} N.~H.,  2006, \mnras, 370, 580

\bibitem[\protect\citeauthoryear{{Levy} \& {Sonett}}{{Levy} \&
  {Sonett}}{1978}]{Levy_Sonett:1978}
{Levy} E.~H.,  {Sonett} C.~P.,  1978, in {T.~Gehrels} ed., IAU Colloq. 52:
  Protostars and Planets {Meteorite magnetism and early solar system magnetic
  fields}.
pp 516--532

\bibitem[\protect\citeauthoryear{{Li}}{{Li}}{1998}]{Li:1998}
{Li} Z.,  1998, \apj, 497, 850

\bibitem[\protect\citeauthoryear{{Ligni{\`e}res}, {Petit}, {B{\"o}hm} \&
  {Auri{\`e}re}}{{Ligni{\`e}res} et~al.}{2009}]{Lignieres:2009}
{Ligni{\`e}res} F.,  {Petit} P.,  {B{\"o}hm} T.,    {Auri{\`e}re} M.,  2009,
  \aap, 500, L41

\bibitem[\protect\citeauthoryear{{Lithwick}}{{Lithwick}}{2009}]{Lithwick:2009}
{Lithwick} Y.,  2009, \apj, 693, 85

\bibitem[\protect\citeauthoryear{{Long}, {Romanova} \& {Lovelace}}{{Long}
  et~al.}{2008}]{Long:2008}
{Long} M.,  {Romanova} M.~M.,    {Lovelace} R.~V.~E.,  2008, \mnras, 386, 1274

\bibitem[\protect\citeauthoryear{{Machida}, {Tomisaka}, {Matsumoto} \&
  {Inutsuka}}{{Machida} et~al.}{2008}]{Machida:2008}
{Machida} M.~N.,  {Tomisaka} K.,  {Matsumoto} T.,    {Inutsuka} S.,  2008,
  \apj, 677, 327

\bibitem[\protect\citeauthoryear{{Maitzen}, {Paunzen} \& {Netopil}}{{Maitzen}
  et~al.}{2008}]{Maitzen:2008}
{Maitzen} H.~M.,  {Paunzen} E.,    {Netopil} M.,  2008, Contributions of the
  Astronomical Observatory Skalnate Pleso, 38, 385

\bibitem[\protect\citeauthoryear{{Marchant}, {Reisenegger} \&
  {Akg{\"u}n}}{{Marchant} et~al.}{2010}]{Marchant:2010}
{Marchant} P.,  {Reisenegger} A.,    {Akg{\"u}n} T.,  2010, ArXiv e-prints

\bibitem[\protect\citeauthoryear{{Markey} \& {Tayler}}{{Markey} \&
  {Tayler}}{1973}]{Markey:1973}
{Markey} P.,  {Tayler} R.~J.,  1973, \mnras, 163, 77

\bibitem[\protect\citeauthoryear{{Markey} \& {Tayler}}{{Markey} \&
  {Tayler}}{1974}]{Markey:1974}
{Markey} P.,  {Tayler} R.~J.,  1974, \mnras, 168, 505

\bibitem[\protect\citeauthoryear{{Mestel}}{{Mestel}}{1970}]{Mestel:1970}
{Mestel} L.,  1970, Memoires of the Societe Royale des Sciences de Liege, 19,
  167

\bibitem[\protect\citeauthoryear{{Mestel} \& {Spitzer} Jr.}{{Mestel} \&
  {Spitzer}}{1956}]{Mestel_Spitzer:1956}
{Mestel} L.,  {Spitzer} Jr. L.,  1956, \mnras, 116, 503

\bibitem[\protect\citeauthoryear{{Morin}, {Donati}, {Petit}, {Delfosse},
  {Forveille} \& {Jardine}}{{Morin} et~al.}{2010}]{Morin_etal:2010}
{Morin} J.,  {Donati} J.,  {Petit} P.,  {Delfosse} X.,  {Forveille} T.,
  {Jardine} M.~M.,  2010, \mnras, 407, 2269

\bibitem[\protect\citeauthoryear{{Mouschovias} \& {Paleologou}}{{Mouschovias}
  \& {Paleologou}}{1979}]{Mouschovias_Paleologou:1979}
{Mouschovias} T.~C.,  {Paleologou} E.~V.,  1979, \apj, 230, 204

\bibitem[\protect\citeauthoryear{{Nakano}}{{Nakano}}{1984}]{Nakano:1984}
{Nakano} T.,  1984, \fcp, 9, 139

\bibitem[\protect\citeauthoryear{{Nordlund} \& {Galsgaard}}{{Nordlund} \&
  {Galsgaard}}{1995}]{Nor_Gal:1995}
{Nordlund} {\AA}.,  {Galsgaard} K.,  1995,
  http://www.astro.ku.dk/~aake/papers/95.ps.gz

\bibitem[\protect\citeauthoryear{{Ouyed}, {Pudritz} \& {Stone}}{{Ouyed}
  et~al.}{1997}]{Ouyed_etal:1997}
{Ouyed} R.,  {Pudritz} R.~E.,    {Stone} J.~M.,  1997, \nat, 385, 409

\bibitem[\protect\citeauthoryear{{Palla} \& {Stahler}}{{Palla} \&
  {Stahler}}{1993}]{Palla_Stahler:1993}
{Palla} F.,  {Stahler} S.~W.,  1993, \apj, 418, 414

\bibitem[\protect\citeauthoryear{{Parker}}{{Parker}}{1975}]{Parker:75}
{Parker} E.~N.,  1975, \apj, 198, 205

\bibitem[\protect\citeauthoryear{{Pessah}}{{Pessah}}{2010}]{Pessah:2010}
{Pessah} M.~E.,  2010, \apj, 716, 1012

\bibitem[\protect\citeauthoryear{{Pessah}, {Chan} \& {Psaltis}}{{Pessah}
  et~al.}{2007}]{Pessah_etal:2007}
{Pessah} M.~E.,  {Chan} C.-k.,    {Psaltis} D.,  2007, \apjl, 668, L51

\bibitem[\protect\citeauthoryear{{Petit}, {Ligni{\`e}res}, {Auri{\`e}re},
  {Wade}, {Alina}, {Ballot}, {B{\"o}hm}, {Jouve}, {Oza}, {Paletou} \&
  {Th{\'e}ado}}{{Petit} et~al.}{2011}]{Petit_etal:2011}
{Petit} P.,  {Ligni{\`e}res} F.,  {Auri{\`e}re} M.,  {Wade} G.~A.,  {Alina} D.,
   {Ballot} J.,  {B{\"o}hm} T.,  {Jouve} L.,  {Oza} A.,  {Paletou} F.,
  {Th{\'e}ado} S.,  2011, \aap, 532, L13

\bibitem[\protect\citeauthoryear{{Pizzolato}, {Maggio}, {Micela}, {Sciortino}
  \& {Ventura}}{{Pizzolato} et~al.}{2003}]{Pizzolato_etal:2003}
{Pizzolato} N.,  {Maggio} A.,  {Micela} G.,  {Sciortino} S.,    {Ventura} P.,
  2003, \aap, 397, 147

\bibitem[\protect\citeauthoryear{{Podio}, {Bacciotti}, {Nisini},
  {Eisl{\"o}ffel}, {Massi}, {Giannini} \& {Ray}}{{Podio}
  et~al.}{2006}]{Podio_etal:2006}
{Podio} L.,  {Bacciotti} F.,  {Nisini} B.,  {Eisl{\"o}ffel} J.,  {Massi} F.,
  {Giannini} T.,    {Ray} T.~P.,  2006, \aap, 456, 189

\bibitem[\protect\citeauthoryear{{Power}}{{Power}}{2007}]{Power:2007}
{Power} J.,  2007, Master's thesis, Queen's University, Kingston, Ontario,
  Canada

\bibitem[\protect\citeauthoryear{{Ray}, {Dougados}, {Bacciotti},
  {Eisl{\"o}ffel} \& {Chrysostomou}}{{Ray} et~al.}{2007}]{Ray_etal:2007}
{Ray} T.,  {Dougados} C.,  {Bacciotti} F.,  {Eisl{\"o}ffel} J.,
  {Chrysostomou} A.,  2007, Protostars and Planets V, pp 231--244

\bibitem[\protect\citeauthoryear{{Rebull}, {Stauffer}, {Megeath}, {Hora} \&
  {Hartmann}}{{Rebull} et~al.}{2006}]{Rebull_etal:2006}
{Rebull} L.~M.,  {Stauffer} J.~R.,  {Megeath} S.~T.,  {Hora} J.~L.,
  {Hartmann} L.,  2006, \apj, 646, 297

\bibitem[\protect\citeauthoryear{{Reisenegger}}{{Reisenegger}}{2009}]{Reiseneg%
ger:2009}
{Reisenegger} A.,  2009, \aap, 499, 557

\bibitem[\protect\citeauthoryear{{Spruit} \& {Uzdensky}}{{Spruit} \&
  {Uzdensky}}{2005}]{Spruit_Uzdensky:2005}
{Spruit} H.~C.,  {Uzdensky} D.~A.,  2005, \apj, 629, 960

\bibitem[\protect\citeauthoryear{{Stahler} \& {Palla}}{{Stahler} \&
  {Palla}}{2005}]{Stahler_Palla:2005}
{Stahler} S.~W.,  {Palla} F.,  2005, {The Formation of Stars}.
Wiley-VCH

\bibitem[\protect\citeauthoryear{{Stassun}, {Mathieu}, {Mazeh} \&
  {Vrba}}{{Stassun} et~al.}{1999}]{Stassun_etal:1999}
{Stassun} K.~G.,  {Mathieu} R.~D.,  {Mazeh} T.,    {Vrba} F.~J.,  1999, \aj,
  117, 2941

\bibitem[\protect\citeauthoryear{{Stassun}, {Mathieu}, {Vrba}, {Mazeh} \&
  {Henden}}{{Stassun} et~al.}{2001}]{Stassun_etal:2001}
{Stassun} K.~G.,  {Mathieu} R.~D.,  {Vrba} F.~J.,  {Mazeh} T.,    {Henden} A.,
  2001, \aj, 121, 1003

\bibitem[\protect\citeauthoryear{{van Ballegooijen}}{{van
  Ballegooijen}}{1989}]{vanBallegooijen:1989}
{van Ballegooijen} A.~A.,  1989, in {G.~Belvedere} ed., Accretion Disks and
  Magnetic Fields in Astrophysics Vol.~156 of Astrophysics and Space Science
  Library, {Magnetic fields in the accretion disks of cataclysmic variables}.
pp 99--106

\bibitem[\protect\citeauthoryear{{Vlemmings}, {Surcis}, {Torstensson} \& {van
  Langevelde}}{{Vlemmings} et~al.}{2010}]{Vlemmings_etal:2010}
{Vlemmings} W.~H.~T.,  {Surcis} G.,  {Torstensson} K.~J.~E.,    {van
  Langevelde} H.~J.,  2010, \mnras, 404, 134

\bibitem[\protect\citeauthoryear{{Vorobyov} \& {Basu}}{{Vorobyov} \&
  {Basu}}{2006}]{Vorobyov_Basu:2006}
{Vorobyov} E.~I.,  {Basu} S.,  2006, \apj, 650, 956

\bibitem[\protect\citeauthoryear{{Wilner} \& {Lay}}{{Wilner} \&
  {Lay}}{2000}]{Wilner_Lay:2000}
{Wilner} D.~J.,  {Lay} O.~P.,  2000, Protostars and Planets IV, pp 509--+

\bibitem[\protect\citeauthoryear{{Woltjer}}{{Woltjer}}{1958}]{Woltjer:1958}
{Woltjer} L.,  1958, Proceedings of the National Academy of Science, 44, 833

\bibitem[\protect\citeauthoryear{{Zhang} \& {Low}}{{Zhang} \&
  {Low}}{2003}]{Zhang_Low:2003}
{Zhang} M.,  {Low} B.~C.,  2003, \apj, 584, 479

\bibitem[\protect\citeauthoryear{{Zhu}, {Hartmann}, {Gammie} \&
  {McKinney}}{{Zhu} et~al.}{2009}]{Zhu_etal:2009}
{Zhu} Z.,  {Hartmann} L.,  {Gammie} C.,    {McKinney} J.~C.,  2009, \apj, 701,
  620

\bibitem[\protect\citeauthoryear{{Zinnecker} \& {Yorke}}{{Zinnecker} \&
  {Yorke}}{2007}]{Zinnecker:2007}
{Zinnecker} H.,  {Yorke} H.~W.,  2007, \araa, 45, 481

\end{thebibliography}

\label{lastpage}

\end{document}